\documentclass[useAMS,usenatbib]{mn2e}
\usepackage{natbib}
\usepackage{color,graphicx} 
\usepackage{amsmath}        
\usepackage{amsfonts}       
\usepackage{amssymb}        
\usepackage{framed}         
\usepackage{upgreek}
\usepackage{xspace}
\usepackage{wrapfig}
\usepackage{rotating}
\newcommand{\lb}{FRB~010724\xspace}

\newcommand{\dmunits}{$\mathrm{cm}^{-3}\,\mathrm{pc}$\xspace}

\def\cite{\citep}

\author[E.~Petroff]{E. Petroff$^{1,2,3}$\thanks{Email:
    epetroff@astro.swin.edu.au}, E. F. Keane$^{1,4,3}$,
  E. D. Barr$^{1,3}$, 
  J. E. Reynolds$^{2}$, J. Sarkissian$^{2}$, \newauthor P. G. Edwards$^{2}$, J. Stevens$^{2}$, C. Brem$^{2}$, A. Jameson$^{1}$,
  S. Burke-Spolaor$^{5}$, S. Johnston$^{2}$, \newauthor N. D. R. Bhat$^{6,3}$, P. Chandra$^{7}$, S. Kudale$^{7}$, S. Bhandari$^{1,3}$ \\ 
  $^{1}$ Centre for Astrophysics and Supercomputing, Swinburne University of Technology, Mail H30, PO Box 218, VIC 3122, Australia. \\ 
  $^{2}$ CSIRO Astronomy \& Space Science, Australia Telescope National Facility, P.O. Box 76, Epping, NSW 1710, Australia \\ 
  $^{3}$ ARC Centre of Excellence for All-sky Astrophysics (CAASTRO). \\ 
  $^{4}$ SKA Organisation, Jodrell Bank Observatory, Cheshire, SK11 9DL, UK \\ 
  $^{5}$ National Radio Astronomy Observatory, 1003 Lopezville Rd., Socorro, NM 87801, USA \\ 
  $^{6}$ International Centre for Radio Astronomy Research, Curtin University, Bentley, WA 6102, Australia \\
  $^{7}$ National Centre for Radio Astrophysics, Tata Institute of Fundamental Research, Pune University Campus, Ganeshkhind, Pune 411 007, India} \date{\today} 

  \title[Identifying the source of perytons]{Identifying the source of perytons at the Parkes radio telescope}

\begin{document}

\maketitle

\begin{abstract}

``Perytons'' are millisecond-duration transients of terrestrial
origin, whose frequency-swept emission mimics the dispersion of an
astrophysical pulse that has propagated through tenuous cold plasma.
In fact, their similarity to \lb\ had previously cast a shadow over
the interpretation of ``fast radio bursts,'' which otherwise appear to
be of extragalactic origin. Until now, the physical origin of the
dispersion-mimicking perytons had remained a mystery. We have
identified strong out-of-band emission at 2.3--2.5 GHz associated with
several peryton events. Subsequent tests revealed that a peryton can be generated at 1.4 GHz when a microwave oven door is 
opened prematurely and the telescope is at an appropriate relative angle. Radio emission escaping from microwave ovens during the magnetron shut-down phase  neatly explain all of the observed properties of the peryton signals. Now that the peryton
source has been identified, we furthermore demonstrate that the
microwaves on site could not have caused \lb. This and other distinct
observational differences show that FRBs are excellent candidates for
genuine extragalactic transients.

\end{abstract}


\section{Introduction}
``Peryton'' is the moniker given to a group of radio signals which
have been reported at the Parkes and Bleien Radio Observatories at
observing frequencies
$\sim1.4$~GHz~\citep{bbe+11,kbb+12,bnm12,sbm14}. The signals are seen
over a wide field-of-view suggesting that they are in the near field
rather than boresight astronomical sources~\citep{kon+14}. They are
transient, lasting $\sim 250$~ms across the band, and the 25 perytons reported in
the literature occured only during office hours and predominantly on
weekdays. These characteristics suggest that the perytons are a form
of human-generated radio frequency interference (RFI). In fact one of
the perytons' defining characteristics --- their wide-field
detectability --- is routinely used to screen out local interference detections
in pulsar searches~\citep{kbb+12,kle+10}.

Perytons' most striking feature, which sets them apart from
`standard' interference signals, is that they are swept in
frequency. The frequency dependent detection of the signal is sufficiently similar to the quadratic form of a bona fide astrophysical signal which has traversed the interstellar medium, that the origin of the first fast radio burst, \lb ~\citep{lbm+07}, was called into question by \citet{bbe+11}. This was mainly based upon the apparent
clustering of peryton dispersion measures (DMs) around $\sim
400\mathrm{pc}\,\mathrm{cm}^{-3}$, which is within $\sim10\%$ of \lb's DM.

Ongoing searches are actively searching for FRBs and perytons and are capable of
making rapidly identifying detections. In this paper we report on three new peryton discoveries from a
single week in January 2015 made with the Parkes radio telescope. In addition to the rapid identification within the Parkes observing band, the RFI environment over a wider frequency range was monitored with dedicated equipment at both the Parkes Observatory and the Australia Telescope Compact Array (located ~400km north of Parkes). For one event, the Giant Metrewave Radio Telescope, in India, was being used to observe the same field as Parkes. Below, in \S~\ref{sec:obs}, we
describe the observing setup and details of the on-site RFI
monitors. In \S~\ref{sec:results} we present the results of the
analysis of our observations, and our successful recreation of peryton
signals. \S~\ref{sec:discussion} discusses, in more depth, the
identified sources of the signals and we compare the perytons to the known FRB population in \S~\ref{sec:perytonFRB}. We present our conclusions in
\S~\ref{sec:conclusions}.

\section{Observations}\label{sec:obs}
As part of the SUrvey for Pulsars and Extragalactic Radio Bursts
(SUPERB\footnote{\textsc{https://sites.google.com/site/publicsuperb/}},
Keane et al. in prep.), at Parkes, real-time pulsar and transient
searches are performed. The live transient searching system developed for SUPERB, which uses the
\textsc{heimdall}\footnote{\textsc{http://sourceforge.net/projects/heimdall-astro/}} single pulse search software package,
is now routinely used by several projects. The survey data are taken using the Berkeley Parkes Swinburne Recorder (BPSR) which is used to produce
Stokes I data from 1024-channel filterbanks covering a total bandwidth of 400 MHz centred at 
1382 MHz with a time resolution of 64 μs and 2-bit digitisation. For each pointing 13 such data streams are recorded, one
for each beam of the multi-beam receiver~\citep{swb+96}. 

The survey has been running since April 2014 to search for pulsars and fast radio bursts.
In December 2014 an RFI monitoring system was installed on the Parkes site identical to ones
which had been in operation at the Australia Telescope Compact Array (ATCA) since November 2014.
The RFI monitor itself is a Rhode \& Schwarz EB500 Monitoring Receiver capable of detecting signals across a wide range of frequencies from 402~MHz to 3~GHz. The frequency and time resolution of the monitoring system are limited to 2 MHz and 10 s, respectively. The antenna is mounted on a rotator, which sweeps out 360 degrees in Azimuth every 12 minutes, then returns to an azimuth of 0 for another 8 minutes before repeating the cycle. A spectrum is produced every 10 seconds, which is obtained by stepping in ~20 MHz steps across the full band. So each 10 sec spectrum has only ~0.1 sec of data at any given frequency. The installation of the monitor gives an unprecedented view of the RFI `environment' at the telescope at any given time and this setup is ideal for identifying very strong signals of RFI which may corrupt observations with the main dish at Parkes. 

In January$-$March 2015 319.2 hours (13.3 days) of 13-beam BPSR data were recorded for the SUPERB survey alone to search for pulsars and fast radio bursts. Total time in the BPSR observing mode in these months was 736.6 hours over a range of observing projects aimed at detecting and studying fast transients. Ultimately 350.7 hours of these observations were searched for perytons in the months of January$-$March 2015 in this work. Three events were discovered, all occurring in the week of 19 January on the 19th (Monday), 22nd (Wednesday), and 23rd (Thursday) of January, 2015 in a rotating radio transient search, the PULSE@Parkes outreach project \cite{hhc09}, and SUPERB, respectively. For the event on the 23rd of January, simultaneous coverage with the Giant Metrewave Radio Telescope (GMRT) was also available, which was shadowing Parkes as part of the SUPERB project's effort to localise FRBs. 

The peryton search for SUPERB and other BPSR data is performed after the Parkes data has been transferred to the gSTAR supercomputer facility at Swinburne University of Technology. The peryton search is performed by summing the frequency-time data of all 13-beams from BPSR and searching these summed data for single pulses using \textsc{heimdall} for pulses with a signal-to-noise ratio (S/N) $\geq$ 10 and DM $\geq$ 10 pc cm$^{-3}$. This method ensures that dispersed pulses occurring in a majority of beams are efficiently detected even if they may be too weak in individual beams to be detected in signle-beam searches. For the perytons identified in January 2015, once the date and UTC time were established the Parkes and ATCA RFI monitor data were checked around the times of the perytons for the presence of signals that might be correlated with the appearance of a peryton at 1.4~GHz. The same search technique was applied to search for perytons in the HTRU intermediate and high latitude surveys of \citet{kjs+10}. The HTRU intermediate latitude survey was conducted between 2008 and 2010 and the high latitude component was conducted between 2009 and 2014. The HTRU survey concluded in February, 2014 and as such no RFI monitor data is available for events detected in these data nor for any peryton detected before those reported here.

\section{Results}\label{sec:results}

\subsection{Three perytons}
\begin{figure*}
  \begin{center}
    \includegraphics[scale=0.25, clip, angle=0]{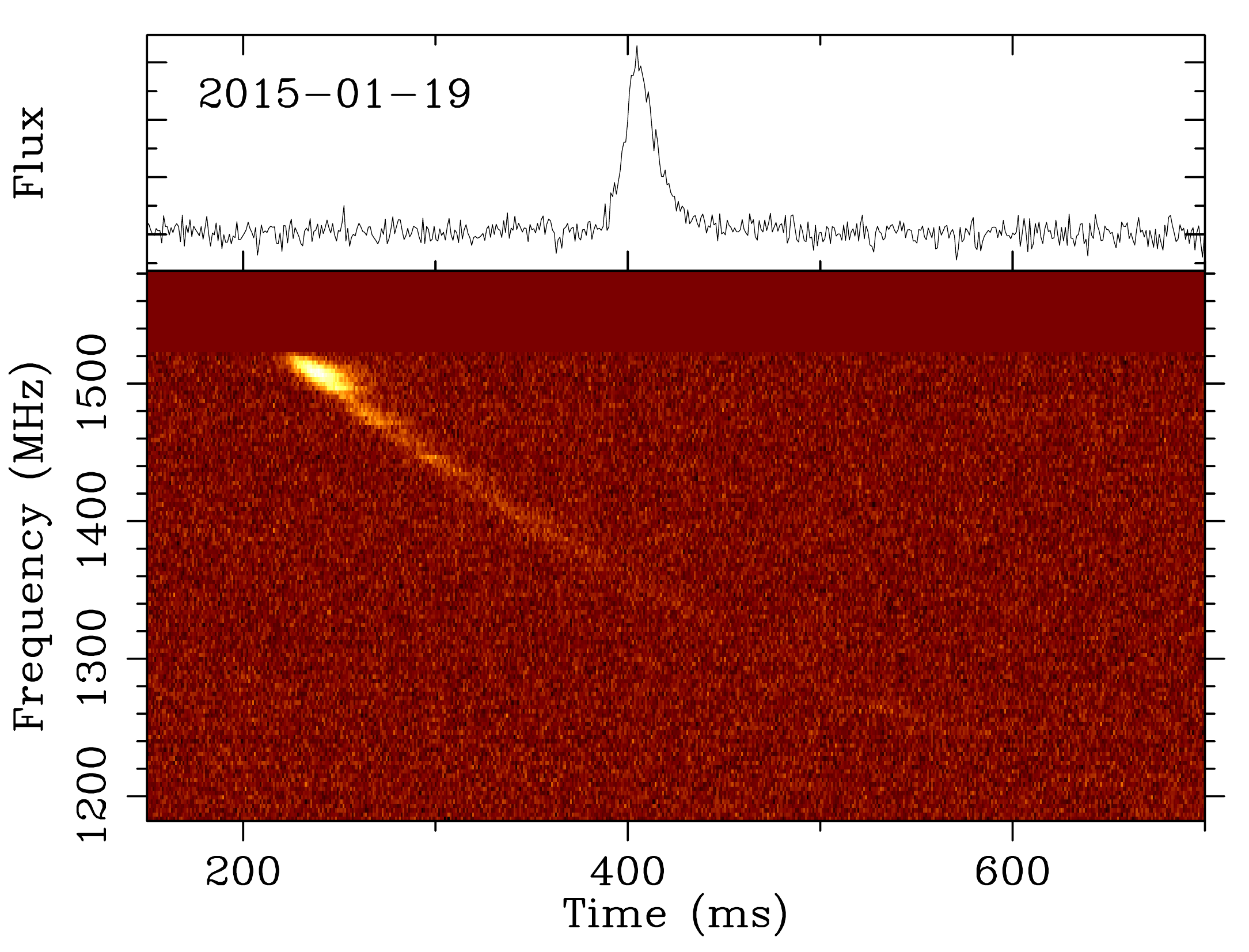}
    \includegraphics[scale=0.25, clip, angle=0]{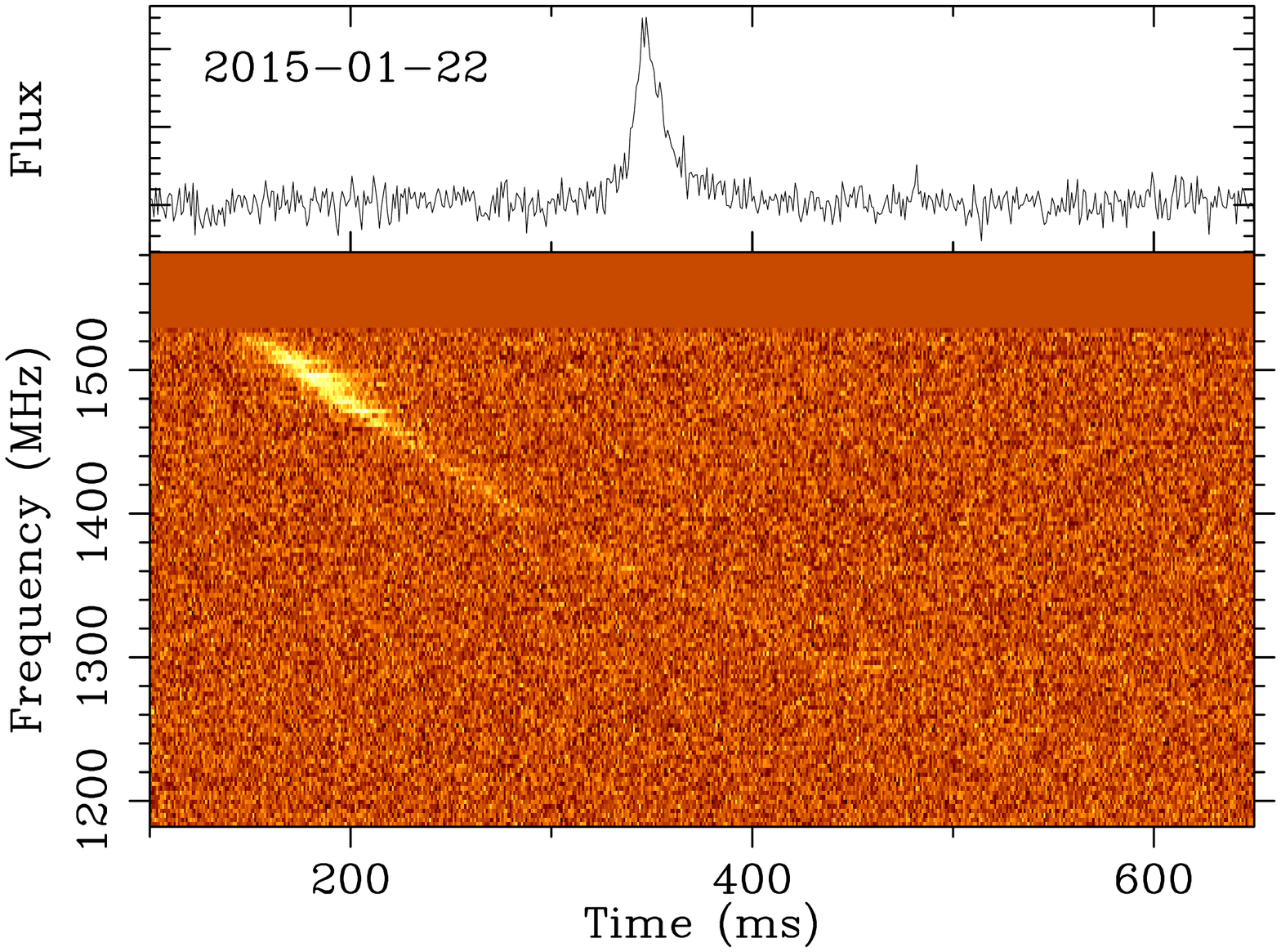}
    \includegraphics[scale=0.25, clip, angle=0]{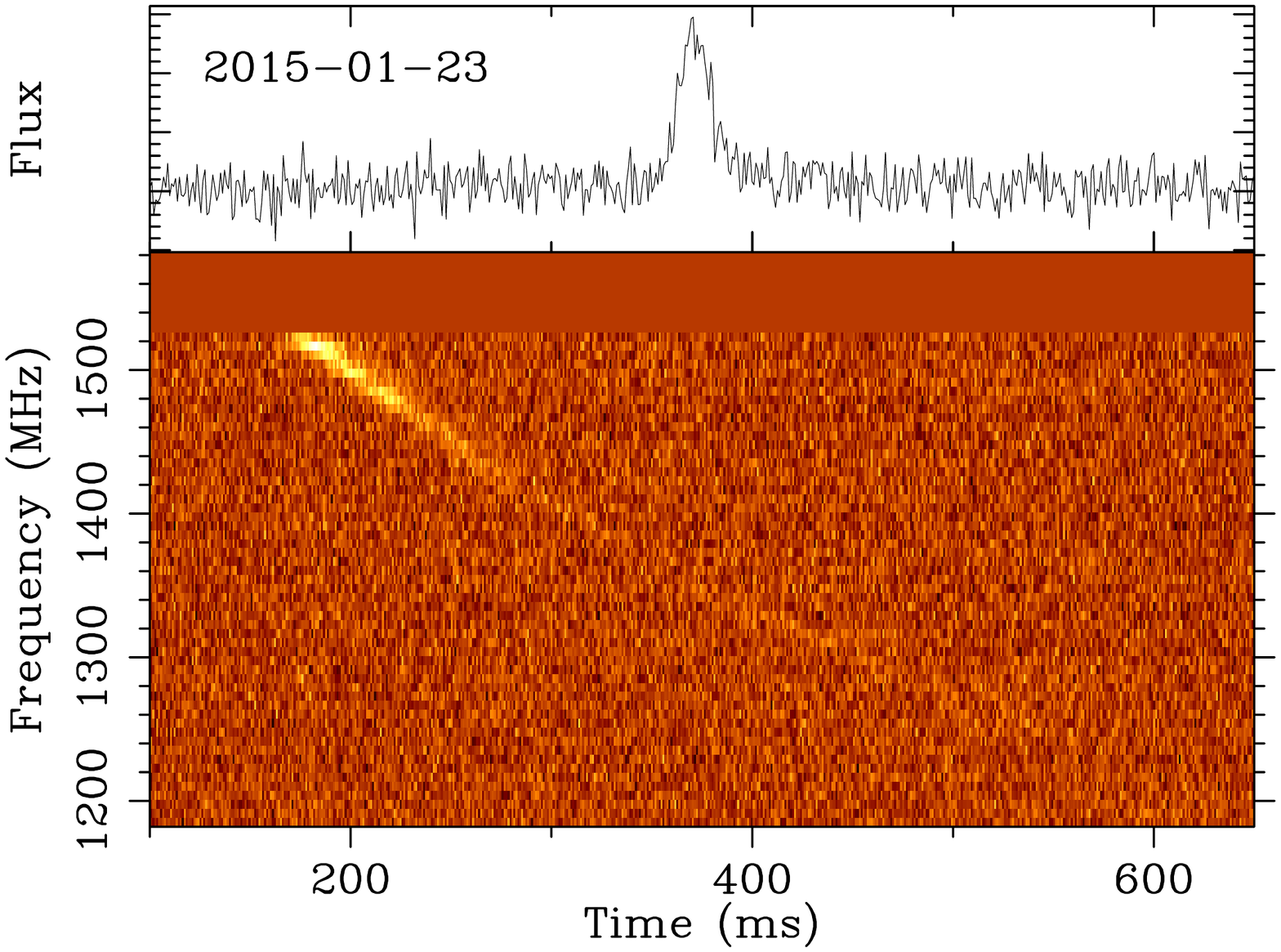}
    \caption{\small{The time-frequency structure of the three January perytons. In the case of events on 2015-01-19 and 2015-01-23 the summed 13-beam data is shown. For 2015-01-22 only beam 01 is plotted as the outer beam data was not recorded to disk.}}\label{fig:sweeps}
  \end{center}
\end{figure*}

The properties of the three perytons discovered in January 2015 are noted in Table~\ref{tab:times} and Figure~\ref{fig:sweeps} shows the
time-frequency structure. These events are typical
perytons in that they are bright and detectable in all beams of the
multi-beam receiver. They are also apparently dispersed or `chirped' in frequency, but not
strictly obeying the quadratic cold plasma dispersion law; signals
from pulsars and FRBs are observed to obey this law precisely \cite{hsh12,tsb+13}. They
have a typical peryton spectrum, being broadband, but brighter at
higher frequencies. Conversely, an off-axis detection of an astronomical source
(i.e. one effectively at infinity) would be suppressed at the highest frequencies,
but the near-field beam pattern is radically different (see
e.g. Figure 10 in \citealt{kon+14}). The existence of a standard template for
peryton spectra and similar DMs also suggests the source, or sources, are at roughly
constant distances and possibly consistently reproducible.

\begin{table*}
  \begin{center}
    \caption{\small{Properties of the perytons from January 2015}} \label{tab:times}
    \begin{tabular}{cccccccc}
      \hline\hline Date  & Time & DM & DM error & S/N  & Width & Telescope  & Telescope  \\
       (dd-mm-yy) & (UTC) & (pc cm$^{-3}$) & & (beam 01) & (ms) & Azimuth (deg) & Elevation (deg) \\
      \hline 
      2015-01-19 & 00:39:05 & 386.6 & 1.7 & 24.8 & 18.5 & 10.7 & 75.3 \\
      2015-01-22 & 00:28:33 & 413.8 & 1.1 & 42.5 & 18.5 & 73.9 & 36.2 \\
      2015-01-23 & 03:48:31 & 407.4 & 1.4 & 10.6 & 18.5 & 323.2 & 40.2 \\
      \hline
    \end{tabular}
  \end{center}
\end{table*}

\begin{figure}
  \begin{center}
    \includegraphics[scale=0.45]{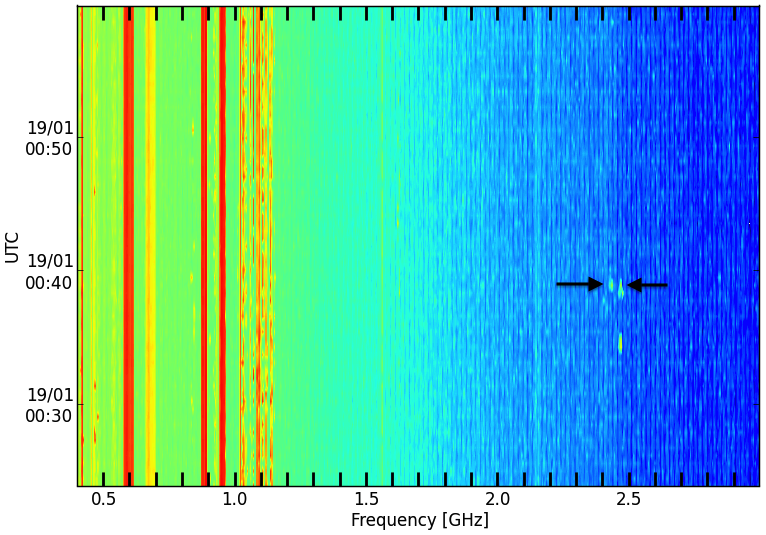}
    \includegraphics[scale=0.45]{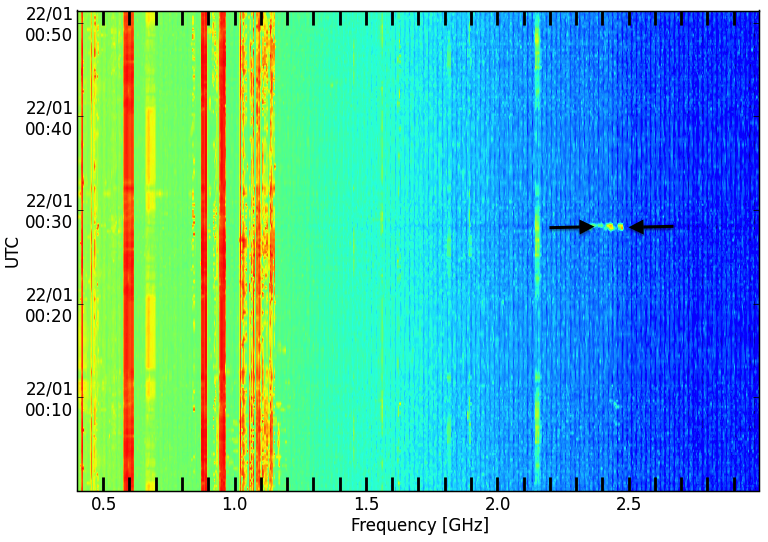}
    \includegraphics[scale=0.45]{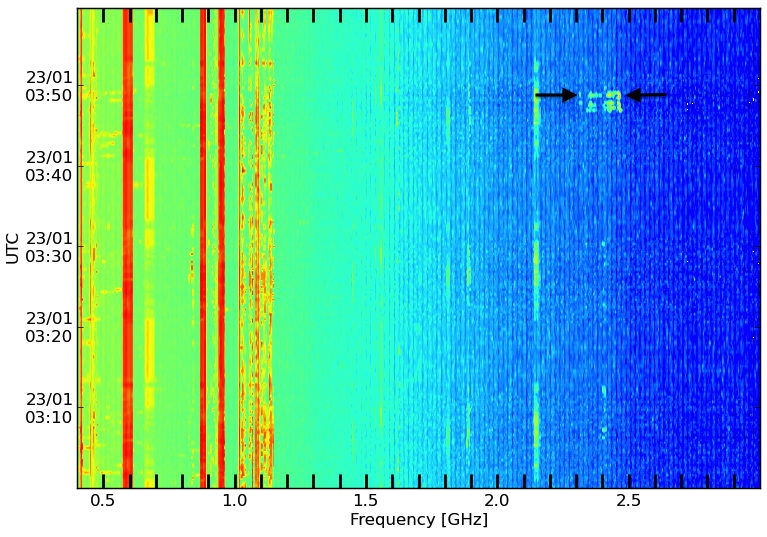}
    \caption{\small{RFI monitor spectra from Parkes for the perytons in the week of 19 January, 2015. The time of peryton has been indicated around the $2.3\sim2.5$~GHz range by black arrows.}}\label{fig:RFI_monitor}
  \end{center}
\end{figure}

\begin{figure*}
\centering
\includegraphics[width=15cm]{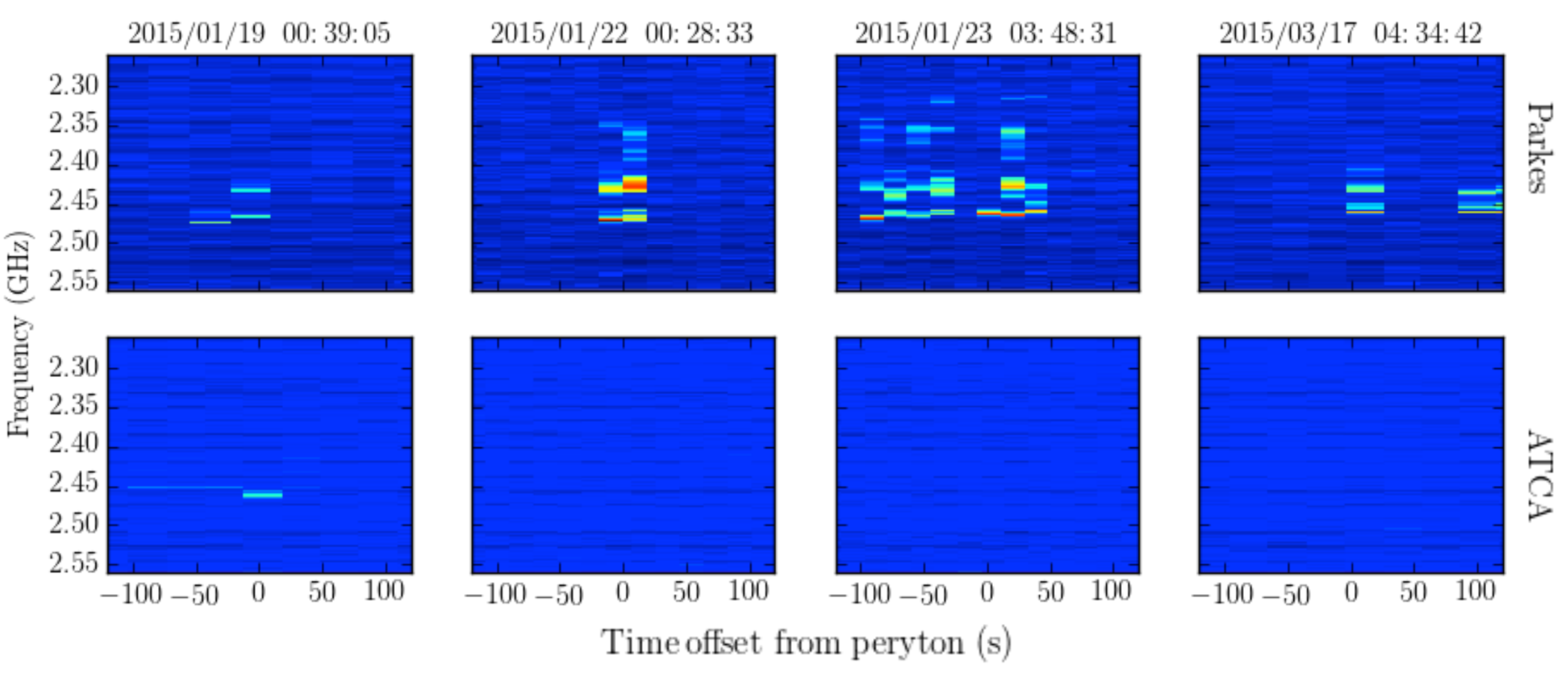}
\caption{\small{RFI monitor data from Parkes and the ATCA between 2.30 and 2.50~GHz around the times of the three January perytons and one peryton from the Woolshed microwave tests (2015-03-17).}}\label{fig:coincidence}
\end{figure*}

These three perytons were the first with simultaneous coverage with additional instruments: the RFI monitors operating at both the Parkes and ATCA sites. For all three events the Parkes RFI monitor detected emission in the frequency range $2.3\sim2.5$~GHz consistent (to less than one time sample) with the time of the 1.4 GHz peryton event. This strongly suggests that the 1.4 GHz millisecond-duration burst is somehow associated with the episodes of 2.4 GHz emission, which last for some 10s of seconds. The broad RFI spectra from the monitor at the times around the perytons is shown in Figure~\ref{fig:RFI_monitor} with the bright emission shown as well as the time of the peryton. Simultaneous emission in the same frequency range was seen in the ATCA data at the time of the first peryton, but no such emission was seen for any other peryton detection, making it likely that this one event was a coincidence, (see Figure~\ref{fig:coincidence}). For the third peryton, simultaneous coverage with GMRT at 325 MHz observing in 2-second snapshots also produced no detection. The detection on only the Parkes site confines the source(s) of the peryton signals to a local origin. 

The 2.3 −- 2.5 GHz range of the spectrum is allocated to ``fixed", ``mobile" and ``broadcasting" uses by
the Australian Communications and Media Authority, and includes use by industrial, scientific and medical
applications, which encompasses microwave ovens, wireless internet, and other electrical items. 
This suggests that the perytons may be associated with equipment operating at $2.3\sim2.5$~GHz, 
but that some intermittent event or malfunctioning, for example, from the equipment's
power supply, is resulting in sporadic emission at 1.4~GHz.

\subsection{Prevalence of $2.3\sim2.5$~GHz signals at Parkes}

\begin{figure}
  \begin{center}
    \includegraphics[scale=0.55]{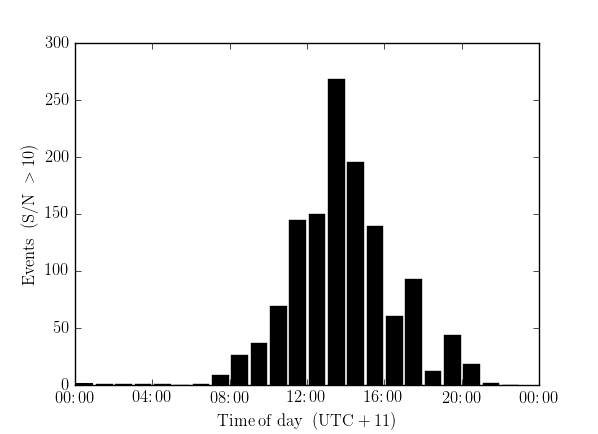}
    \caption{\small{Number of narrow-emission spikes detected with the RFI monitor with S/N $>$ 10 in a 60 MHz window around 2.466~GHz between 18 January and 12 March, 2015.}}\label{fig:burst_hist}
  \end{center}
\end{figure}

As can be seen in Figure 2 there is at least one case where a single peryton is detected but there are multiple or ongoing detections at $2.3\sim2.5$~GHz around the time of the peryton. This already indicates that while peryton detections at 1.4 GHz coincide with episodes of emission at higher frequency, the higher frequency emission can occur without generating a peryton. More detailed inspection of the archival RFI monitor data at Parkes gives an indication of the prevalence of these episodes at higher frequencies. In the months investigated several hundred spikes of emission are detected in the frequency range $2.3\sim2.5$~GHz. These events cluster in time of day and are much more common during daytime (between the hours of 9am and 5pm local time). A time-of-day histogram of these spikes over the period of 18 January to 12 March, 2015 is plotted in Figure~\ref{fig:burst_hist}. This is entirely consistent with the use of microwave ovens and other electrical equipment. Tests at Parkes confirmed that microwave ovens produced detectable levels of $\sim$2.4~GHz emission in the RFI monitoring equipment independent of the azimuth of the rotator. Standard practice at ATNF observatories is not to allow the use of microwave ovens on site when observing in the 2.4~GHz band is taking place.

\subsection{Archival perytons}

Using the search technique described in \S~\ref{sec:obs} 15 perytons were found in the HTRU intermediate latitude survey and an additional 6 perytons were found in a search of 75\% of the high latitude survey. While the RFI monitor had not yet been set up on site and the RFI environment is impossible to recover, we can use these perytons to study the ensemble properties. Combining the perytons from January 2015, HTRU, \citet{bbe+11}, \citet{kbb+12}, and \cite{bnm12} the total number of perytons is 46. The properties of these sources, especially how they relate to the population properties of FRBs is discussed in more detail in \S~\ref{sec:perytonFRB}.

\subsection{Generating perytons}
With the recognition that peryton signals are likely to be associated with equipment emitting
at 2.3$\sim$2.5 GHz, an effort was made to try to identify such equipment on site, 
 and attempt to `create' a peryton. As microwave ovens are known to emit in this frequency range and the could potentially produce short-lived emission the site microwaves were the focus of our initial tests for reproducing peryton signals.

There are three microwaves on site in close proximity to the telescope that experience frequent use located in the tower below the telescope, the visitors centre and the staff kitchen located in the building traditionally referred to as the Woolshed. There are two additional microwaves at the observers quarters approximately 1 km from the main site. The first tests occurred on 27 February, 2015 during scheduled maintenance while the telescope was stowed at zenith. The BPSR system was turned on for all 13 beams and the three microwaves on site were run on high and low power for durations of 10 -- 60 s. In each test the load in the microwave was a ceramic mug full of water. In the first set of tests a single peryton was detected during tests of the tower microwave with a DM of 345 pc cm$^{-3}$. The detection of radiation from the tower microwave would be very surprising as the tower is shielded on the windows and in the walls and the dish surface blocks the line of sight to the receiver in the cabin at the prime focus. However it was later determined that the Woolshed microwave was also in use at the time, 
unrelated to these tests, and might potentially have been the source of the peryton.

The second set of tests were conducted on 12 March, 2015, this time pointing the telescope at azimuth and elevation combinations where we often see perytons. From the 21 perytons discovered in the HTRU survey and the known pointing locations a broad estimate of the peryton rate as a function of azimuth and elevation can be calculated. For the HTRU perytons the rate is highest at an azimuth and elevation of ($\sim$130$^\circ$, 65$^\circ$) and when pointing near zenith. An initial test was conducted with the microwaves while pointing the telescope at these locations and no perytons were seen.

\begin{figure}
  \begin{center}
    \includegraphics[width=9cm]{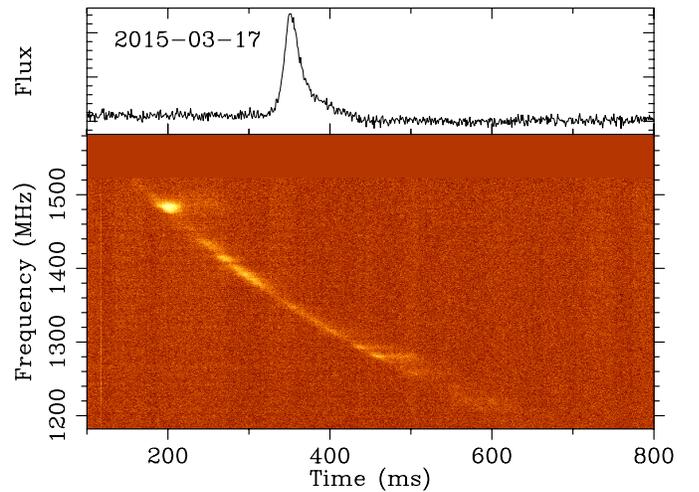}
    \caption{\small{One of the bright perytons generated during the test on 17 March with DM = 410.3 \dmunits. RFI monitor data at the time of this peryton is shown in Figure~\ref{fig:coincidence}.}}\label{fig:testperyton}
  \end{center}
\end{figure}

\begin{figure}
\begin{center}
\includegraphics[width=8cm]{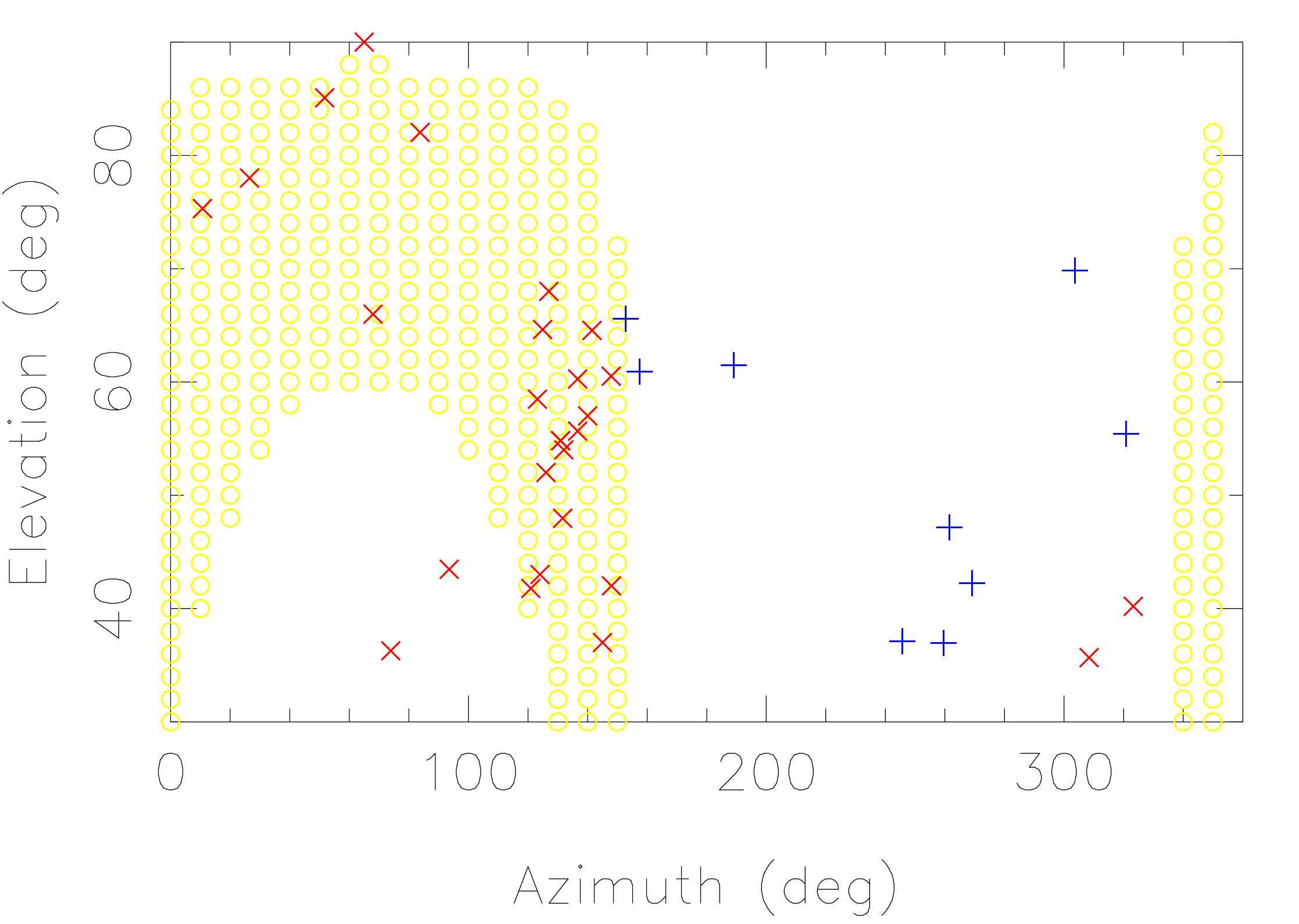}
\caption{\small{Azimuth and elevation combinations for which there is a direct line of sight from the microwave oven in the Woolshed to the multibeam receiver are broadly shown with circles. The pointing directions for the detected perytons with DMs $\sim400$ pc cm$^{-3}$ (crosses) and $\sim200$ pc cm$^{-3}$} (pluses) are also shown.}\label{fig:lineofsight}
\end{center}
\end{figure}

The decisive test occurred on 17 March, 2015 when the tests were repeated with the same microwave setup but instead of waiting for the microwave cycle to finish the microwave was stopped by opening the door. This test produced 3 bright perytons from the staff kitchen microwave all at the exact times of opening the microwave oven door with DMs of 410.3, 410.3, and 399.6 pc cm$^{-3}$, (see Figures~\ref{fig:coincidence} and~\ref{fig:testperyton}). With knowledge that this mode of operation of a microwave oven could produce perytons, we examined the range of azimuths and elevations at which there was direct line of sight from the microwave to the multibeam receiver (i.e., the underside of the focus cabin). As is apparent in Figure~\ref{fig:lineofsight}, almost all the perytons with DMs $>$ 300 occurred when there was visibility of the focus cabin from the Woolshed microwave. This left the smaller sample of perytons with lower DMs, which were, however, consistent with an origin at the visitors centre or the Quarters. (This sample also included all five events which had been detected on the weekend, when there were generally no staff on site and the Woolshed not in use.) Similar tests were performed with a previously installed microwave oven in the visitors centre and 6 perytons were seen at the times corresponding to opening the door, however these perytons had DMs of 206.7, 204.9, 217.0, 259.2, 189.8, and 195.2 pc cm$^{-3}$. This process does not generate a peryton every time, however; in fact perytons appear to be generated with a $\sim$50\% success rate.

A bimodal distribution of peryton DMs can be accounted from at least two microwaves on site being used and stopped in this manner. The detectability of perytons with a given DM from a microwave stopped this way depends on the direction in which the telescope is pointing. The receiver is sensitive to perytons when the microwave oven producing the bursts has a direct line of sight to the focus cabin and receiver of the telescope, i.e., a line of sight not blocked
by the surface of the telescope, yet still seeing the underside of the focus cabin. As shown in Figure~\ref{fig:lineofsight},
for the Woolshed (located 100m from the Dish at an azimuth of 65$^\circ$), the broadest range of elevations
providing a direct line of sight are offset by $\sim$80$^\circ$ in azimuth.

\subsection{The Peryton Cluster of 23rd June 1998}
Of the 46 perytons detected at Parkes since 1998 some 16, more than a third of the total, occurred within a period of just seven minutes, on 23rd June 1998. All have a DM consistent with an origin in the Woolshed. Kocz et al. (2012) noted that the interval between consecutive events is clustered around 22 seconds. In this more complete sample we find that indeed eight of the 15 intervals between consecutive events fall within the range 22.0+/-0.3 seconds, which is exceedingly unlikely to have been produced by manually opening the oven. Rather, we believe that the operator had selected a power level of less than 100\%, causing the magnetron power to cycle on and off on a 22-second cycle, the period specified in the manufacturer's service manual and confirmed by measurement. It appears likely that over this seven-minute period the oven produced a peryton on all or most completions of this 22-second cycle but that the operator stopped the oven manually several times by opening the door, each time restarting the 22-second cycle.

Kocz et al. also noted a clustering of event times modulo 2 seconds (their Figure 2). This can be explained if the 22-second cycle is derived from a stable quartz crystal oscillator, which is almost certainly the case as the oven has a digital clock display.

However we have been unable to repeat the production of perytons in this manner. The principal difficulty is to account for the peryton energy escaping the oven's shielded enclosure without opening the door. A transitory fault condition seems an unlikely possibility, given the oven has continued to operate reliably for a further 17 years. We conjecture that on this occasion the operator inadvertently compromised the shielding by placing conducting material in the oven, perhaps Aluminium cooking foil that became caught between the door and the body of the oven, creating a unintended antenna, but we have yet to devise an acceptable test of this scenario.


\section{Discussion}\label{sec:discussion}
The two ovens responsible for most or all of the observed perytons are from the same manufacturer (Matsushita/National) and are both in excess of 27 years of age though still working reliably. Our tests point clearly to the magnetron itself as the source of the perytons since these are not detected unless the oven door is opened. Further, our analysis of the peryton cluster of 23rd June 1998 implies the perytons are a transient phenomenon that occurs only when the magnetron is switched off. That we have observed perytons from at least two ovens over 17 years suggests that they are not the product of an unusual failure or fault but are inherent to, and long-lived in, at least some common types of oven. The magetron used in the Woolshed oven (type 2M210-M1) was used by Matsushita in new microwave ovens for at least a decade and remains readily available.

However the physical process that generates the swept or `chirped' emission that defines these perytons is obscure. The duration of the perytons is also a puzzle. The Woolshed oven has a simple HV supply comprising a 2kVAC mains step-up transformer and Villard voltage doubler/rectifier, with no additional filtering. The magetron supply voltage should decay rapidly after switch-off over a few mains cycles (of 20 ms) but the perytons have typical durations of 250 ms or more, decaying in power by only a factor of 3 or so over this time (e.g. Figure 3. of \citeauthor{bnm12} \citeyear{bnm12}).


By nature, magnetrons are highly non-linear devices and the mode competition occurring at the start-up and shut-down of the microwave can cause excitation within the magnetron. Magnetron cavities have several spacings through which electrons flow. Over time the edges of these cavities may become worn down and arcing may occur across these cavities during start-up and shut-down. This arcing may produce a spark observable at other frequencies than those intended in the microwave specifications. The microwave itself should act as a Faraday cage and block these signals from exiting the microwave oven cavity. However, opening the door of the microwave during shut-down would allow for these signals to propagate externally. Escaping sparks at 1.4~GHz could be the perytons we see with the receiver \cite{ys95,bss07}.



\section{Relevance to FRBs}\label{sec:perytonFRB}

\subsection{Differences in observed properties}
Having originally cast doubt on the first FRB discovered, \lb, the
origin of perytons has since cast a shadow on the interpretation of
FRBs as genuine astrophysical pulses. We therefore wish to explicitly
address whether perytons and FRBs could have a common origin.
Even with the source of perytons identified as on-site RFI the
question may remain as to whether the progenitors of FRBs and perytons
are related or even the same event at different distances.
Fundamental aspects of the FRB and peryton populations differ. The
distribution of perytons in time of day occurrence and DM is highly
clustered and very strongly indicative of a human-generated signal.
The DM and time of day detections of perytons and FRBs are compared in
Figure~\ref{fig:FRB_peryton_comparison}. In the case of the perytons
the clustering around the lunchtime hour becomes even more pronounced
once this AEDT correction is applied. The FRB distribution in time of
day is consistent with a random distribution, which would be observed
as essentially flat perhaps with a slight dip in number during office
hours where occasional telescope maintenance is performed. 

Similarly, the bimodal DM distribution of the peryton population can be clearly
seen in the larger peryton sample. No clear DM clustering can yet be
identified for the FRBs although such a distribution may become clear
with a population of 1000s of sources if FRBs are cosmological
\citep{mcq14,mkg15}.
Finally, a microwave oven origin is generally not well suited to
explaining other observed properties of FRBs, such as the clear
asymmetric scattering tails observed in some FRBs, the consistency
with Komolgorov scattering \citep{tsb+13}, and the apparent deficit of
detections at low Galactic latitudes. These are major indicators of a genuine
astrophysical population \citep{psj+14,bb14}.

\subsection{What is \lb?}
\begin{figure*}
  \begin{center}
    \includegraphics[scale=0.4,trim = 0mm 0mm 0mm 0mm, clip, angle=0]{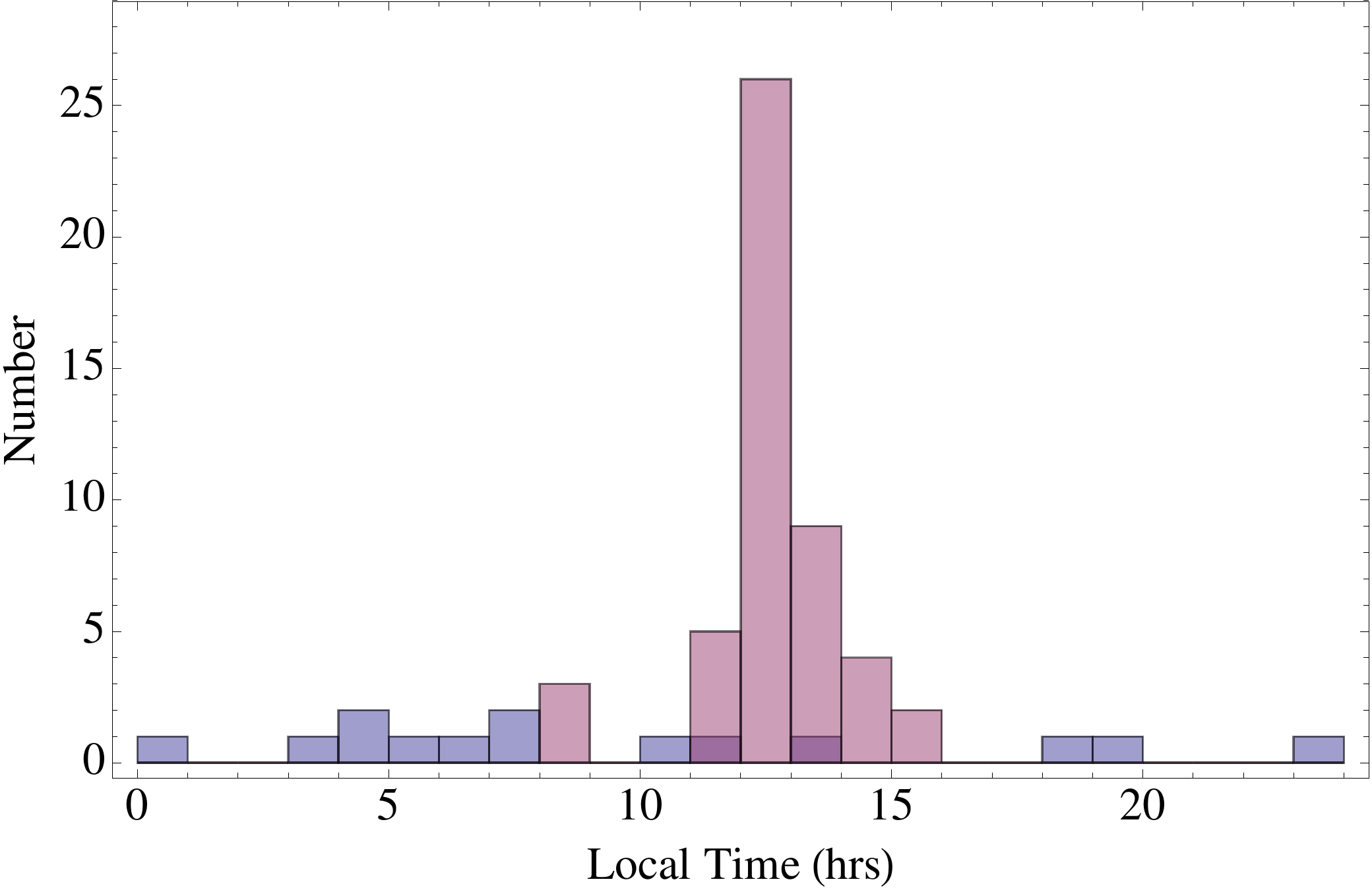}
    \includegraphics[scale=0.4,trim = 0mm 0mm 0mm 0mm, clip, angle=0]{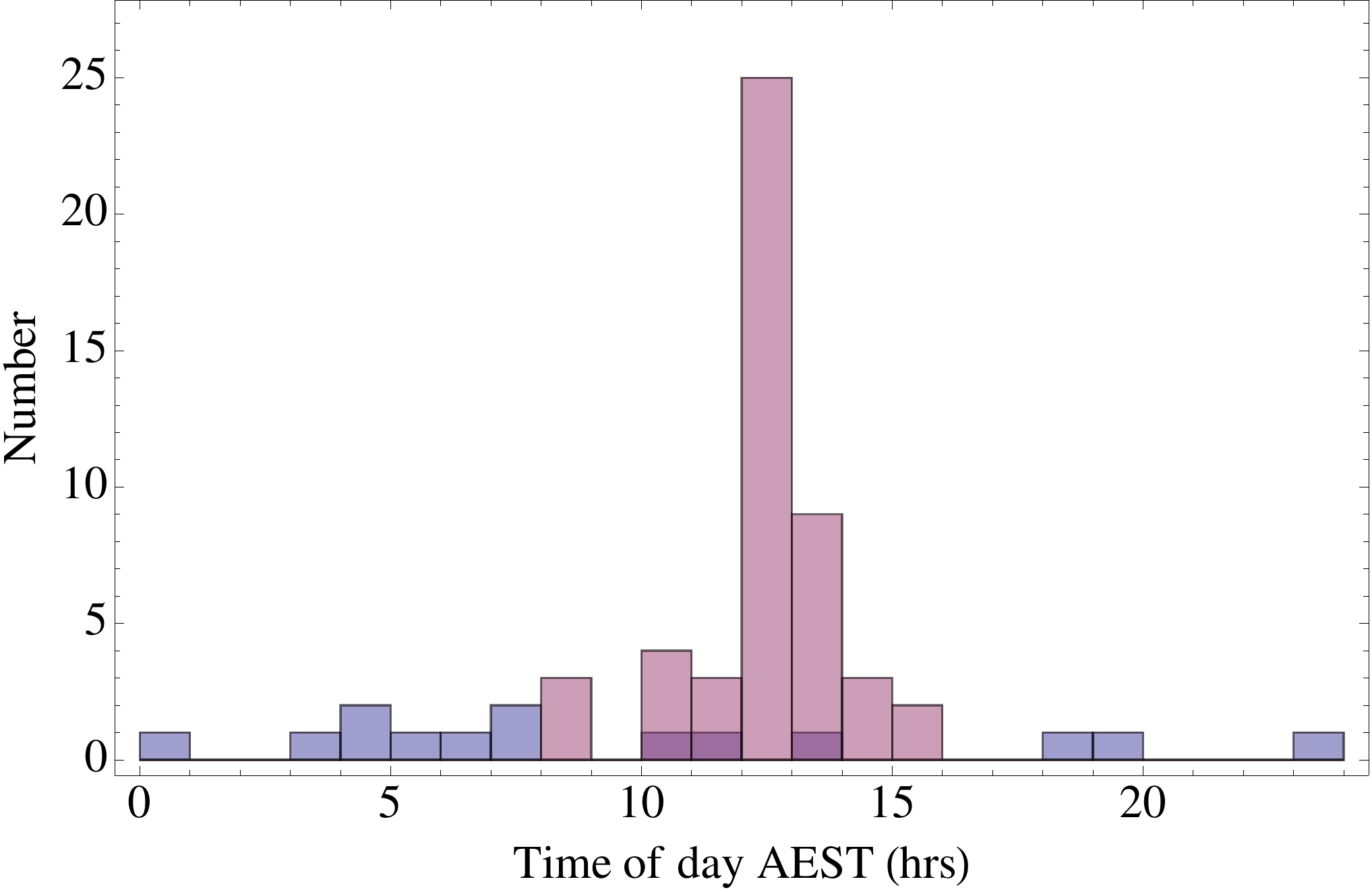}
    \includegraphics[scale=0.4,trim = 0mm 0mm 0mm 0mm, clip, angle=0]{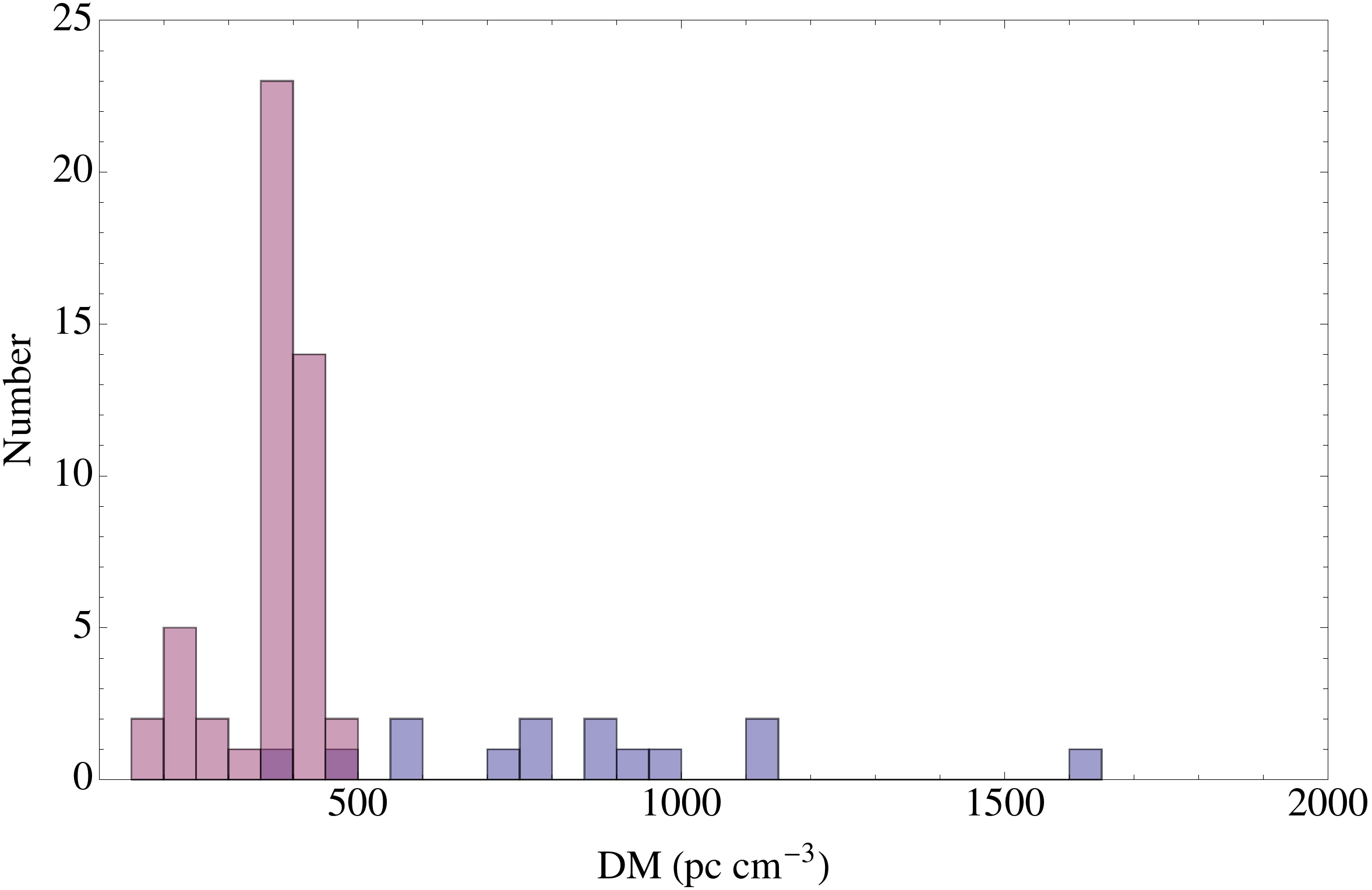}
    \caption{\small{The FRB (blue) and peryton (red)
        distributions as a function of the local time with Australian
        Eastern Daylight Savings Time accounted for (top left), time in Australian Eastern Standard Time (top right) and as a function of DM over the entire range searched (bottom). Clearly the FRB distribution is uniform throughout the
        day, whereas the peryton signals peak strongly during office
        hours (particularly around lunch time). A random distribution would look approximately flat,
        with a slight dip during office hours where occasional
        maintenance is carried out. The bi-modal peryton distribution with peaks at $\sim200$ and
        $\sim400\;\mathrm{cm}^{-3}\,\mathrm{pc}$ is evident.}}\label{fig:FRB_peryton_comparison}
  \end{center}
\end{figure*}

With an understanding of the conditions under which perytons are
generated, we can reconsider the ``Lorimer Burst,'' \lb
\citep{lbm+07}. As noted by \citet{bbe+11} and as is evident in Figure
\ref{fig:FRB_peryton_comparison}, the DM of 375\,\dmunits for this
burst is entirely consistent with the DM$\sim$400\,\dmunits events we
now refer to as Woolshed perytons. However, there are critical
differences. The bright detection in 3 beams is indicative of a
boresight detection. Furthermore, the event occurred with the
telescope pointing almost due south, and the line of sight from the
Woolshed microwave to the focus cabin is completely blocked by the
telescope surface. While there is line of sight visibility from the
visitors centre at this time, the DM is not consistent with the
visitors centre microwave. Additionally, the event occurred at 19:50
UT, i.e., 5:50 am AEST, when the visitors centre is closed and
unstaffed. We conclude the evidence in favour of \lb\ being a genuine
FRB is strong.

\subsection{Deciphering new transient events}
To discern between new millisecond transient detections, this work has
demonstrated two critical discriminants that divide FRBs and perytons.
A common, known RFI emission from microwaves---as detected
concurrently to all perytons presented here---is at $2.3-2.5$~GHz.
Thus, an FRB detected with a non-detection of any $2.3-2.5$~GHz, which
we propose as a key characteristic of the Parkes perytons---would be
another nail in the coffin for any association. It should be noted
that while there are $2.3-2.5$~GHz events with no L-band detection,
there are not the converse, so there is some statistical probability
that an FRB occurs by chance around the same time as an FRB,
particularly if it is detected during daytime (Fig
\ref{fig:burst_hist}).

Second, as with \lb, given that the telescope cannot point directly at
a microwave, fabricating a detection that does not appear in all
beams, our results show that perytons can be discerned from FRBs by
using a multibeam system to identify sky-localized events. 
For an event to appear point-like within the
multibeam receiver's beam pattern, as FRBs do, the target must be in
the Fraunhofer regime.

\section{Conclusions}\label{sec:conclusions}

Three peryton detections were made at the Parkes radio telescope on
three separate days during the week of 19 January, 2015. The
installation of a new broadband RFI monitor allowed for the first
correlation between the peryton events and strong out of band emission
at 2.3--2.5 GHz of local origin. Additional tests at Parkes revealed
that peryton events can be generated under the right set of conditions
with on-site microwave ovens and the behaviour of multiple microwaves
on site can account for the bimodal DM distribution of the known
perytons. Peryton searches in archival survey data also allowed for
the detection of a further 21 bursts from the HTRU survey alone. 
A comparison of the population properties of FRBs and perytons revealed
several critical conclusions:
\begin{itemize}
\item Perytons are strongly clustered in DM and time of day, strongly
indicative of man-made origins, whereas FRBs are not.
\item FRB detections to date faithfully follow cold plasma dispersion;
some have shown clear scattering tails whose frequency-dependent width
follows a Kolmogorov spectrum; FRBs appear to avoid the Galactic
plane. Perytons do not exhibit these properties.
\item The peryton-causing ovens on Parkes site could not have produced
\lb, indicating that this burst is in fact an FRB rather than a
peryton.
\item A direct test of ``peryton vs.~FRB'' can be made via the
detection or non-detection, respectively, of concurrent 2.3--2.5 GHz
emission.
\end{itemize}
We have thus demonstrated through strong evidence that perytons and
FRBs arise from disparate origins. There is furthermore strong
evidence that FRBs are in fact of astronomical origin.



\section*{Acknowledgements}\label{sec:public_data}
We thank the team from the PULSE@Parkes project for use of the beam 01 data from their observing session. EP would like to thank I. Morrison for useful discussion and support. Following our realization that microwave ovens were the likely source of the perytons we had useful discussions about the functioning of the devices with J. Benford on March 13. 
This work used the gSTAR national facility which is funded by
Swinburne and the Australian Government’s Education Investment
Fund. 
The Parkes Radio Telescope is part of the Australia Telescope
National Facility, which is funded by the Commonwealth of Australia
for operation as a National Facility managed by CSIRO. 
EP, EDB, and EFK acknowledge the support of the Australian Research Council Centre
of Excellence for All-sky Astrophysics (CAASTRO), through project
number CE110001020.

\bibliography{journals,journals_apj,psrrefs,modrefs,crossrefs}
\bibliographystyle{mnras}

\end{document}